\documentclass[%
 reprint, 
 amsmath,amssymb,
 aps, physrev,
]{revtex4-2}

\usepackage{graphicx}% Include figure files
\usepackage{dcolumn}% Align table columns on decimal point
\usepackage{bm}% bold math
\usepackage{braket}
\usepackage{hyperref}% add hypertext capabilities
\begin{document}

\preprint{APS/123-QED}

\title{Quantum spin-heat engine with trapped ions} % 

\author{Andr\'e R. R. Carvalho}
\author{Liam J. McClelland}%
 \email{Contact author: l.mcclelland@griffith.edu.au}
\author{Erik W. Streed}
\author{Joan Vaccaro}
\affiliation{%
 Queensland Quantum and Advanced Technologies Research Institute, Griffith University, Brisbane, Queensland 4111 Australia}%

\date{\today}% It is always \today, today,

\begin{abstract}
We propose an ion-trap implementation of the Vaccaro, Barnett and Wright et al. spin-heat engine (SHE); a hypothetical engine that operates between energy and spin thermal reservoirs rather than two energy reservoirs. The SHE operates in two steps: first, in the work extraction stage, heat from a thermal energy reservoir is converted into optical work via a two photon Raman transition resonant with close-to energy degenerate spin states; second, the internal spin states are brought back to their initial state via non-energetic information erasure using a spin reservoir. The latter incurs no energy cost, but rather the reset occurs at the cost of angular momentum from a spin bath that acts as the thermal spin reservoir. The SHE represents an important first step toward demonstrating heat engines that operate beyond the conventional paradigm of requiring two thermal reservoirs, paving the way to harness quantum coherence in arbitrary conserved quantities via similar machines.
\end{abstract}

\maketitle

%\tableofcontents

\section{Summary}\label{sec: Summary}
Thermodynamical heat engines produce useful work by converting heat contained within a higher temperature reservoir as it thermalizes with a lower temperature reservoir. In this conversion, only part of the heat extracted is transformed into work, with the engine’s efficiency ultimately limited by Carnot’s theorem. From a purely information-theoretical point of view, energy doesn’t need to play a preferred role in statistical mechanics, as the principle of maximum entropy does not depend on the particular physical quantity being considered. This concept led to a formulation of generalised statistical mechanics using multiple conserved quantities \cite{jaynes1957information, jaynes1957information2, janes1982rationale}.

Inspired by Jaynes, Vaccaro and Barnett (VB) showed that a memory device can be erased without energy cost, as long as the price is paid in another conserved quantity \cite{Vaccaro_2011}. This was a re-interpretation of Landauer's principle, with the heat dissipated during the information erasure now replaced by the angular momentum equivalent, spintherm.  This opened a way to interpret Maxwell’s demon thought experiment as a way to extract work from a single thermal reservoir at the cost of angular momentum, for example. A bath of spin polarized atoms became the effective memory of the demon, whose memory can be erased with resources other than energy. This idea was applied to Quantum dots (named the QDSHE) by Wright et al. \cite{wright2018quantum}, while the concepts of the spin equivalent of work and heat (called spinlabour $\mathcal{L}$ and spintherm $\mathcal{Q}$ respectively) were applied to information erasure in discrete fluctuating systems by Toshio et al. \cite{croucher2017discrete}. 

Several interesting studies have been done in recent years by implementing generalised Gibbs distribution `a la Jaynes \cite{Pozsgay_2017, PhysRevE.101.042117, Langen_2016, Guryanova_2016, mcclelland2025beyond}. In the context of global quantum quenches in XXZ Heisenberg spin chains, Pozsgay et al. \cite{Pozsgay_2017} showed theoretically that the generalized Gibbs ensemble can be implemented within the quantum transfer matrix method. The thermal state of subsystems exchanging noncommuting conserved quantities are termed ‘the non-Abelian thermal state’ (NATS) by Halpern et al. \cite{PhysRevE.101.042117}. This thermal state has a generalized Gibbs form reinforcing Jaynes’s derivation of thermodynamics to arbitrary conserved quantities through the principle of maximum entropy. The derivation of NATS from three various (microcanonical derivation, dynamical considerations and resource theory) arguments provide tools applicable to quantum noncommutation in thermodynamics. Langen et al. \cite{Langen_2016} showed both theoretical and experimental observation of a degenerate one-dimensional Bose gas relaxing to a state described by generalised Gibbs ensemble that exhibit nontrivial conserved quantities.

An interesting work close to the context of our manuscript was done by Guryanova et al. \cite{Guryanova_2016} that deals with the generalization of thermodynamics with multiple conserved quantities, ‘where energy may not even be part of the story’. For a standard thermodynamic scenario consisting of a thermal bath, an external system out of equilibrium with respect to the bath and a number of batteries, the extracted work to be stored in batteries are limited by the change in Gibbs free energy of the system. Instead of recovering the generalised thermal state at a cost of generalised Gibbs free energy they coined the term `free entropy' - a dimensionless quantity defined with respect to a set of inverse temperatures. The state that minimizes the free entropy for fixed inverse temperatures gives the generalised thermal state. They showed that the change in free entropy puts the limit on the combination of conserved quantities that can be extracted. The results hold both for commuting and non-commuting observables.

These concepts have since been applied by McClelland \cite{mcclelland2025beyond} to a hypothetical battery that operates on the entropic differences between reservoirs of arbitrary conserved quantities, which he called the entropy battery. He takes a statistical mechanics perspective, generalizing some of the results obtained in this manuscript to ensembles with arbitrary discrete states per particle per degree of freedom. He focused on a cold thermal energy and spin polarized reservoir to extract energetic work from a hot thermal bath.  With $d\ge2$ internal spin states, and initial temperatures ranging between room temperature and the operating temperature of typical power plants, such a device could surpass the standard Carnot efficiency limit. He utilized unitary heat engines that operate adiabatically across multiple conserved quantities such as the one discussed in this manuscript. 

We use these ideas to show it is possible to build an engine similar to the Write et al. QDSHE and the Vaccaro-Barnett spin heat engine (SHE) in a trapped ion system that functions between a thermal energy and a spin reservoir. The engine cycle involves exchanges between heat and work in terms of energy, but also of the equivalent quantities for spin angular momentum ($\mathcal{Q}$ and $\mathcal{L}$). As a consequence of the working mechanism, there is no fundamental limit in the amount of work that can be extracted from heat. 

\begin{figure}
    \centering
    \includegraphics[width=0.8\linewidth]{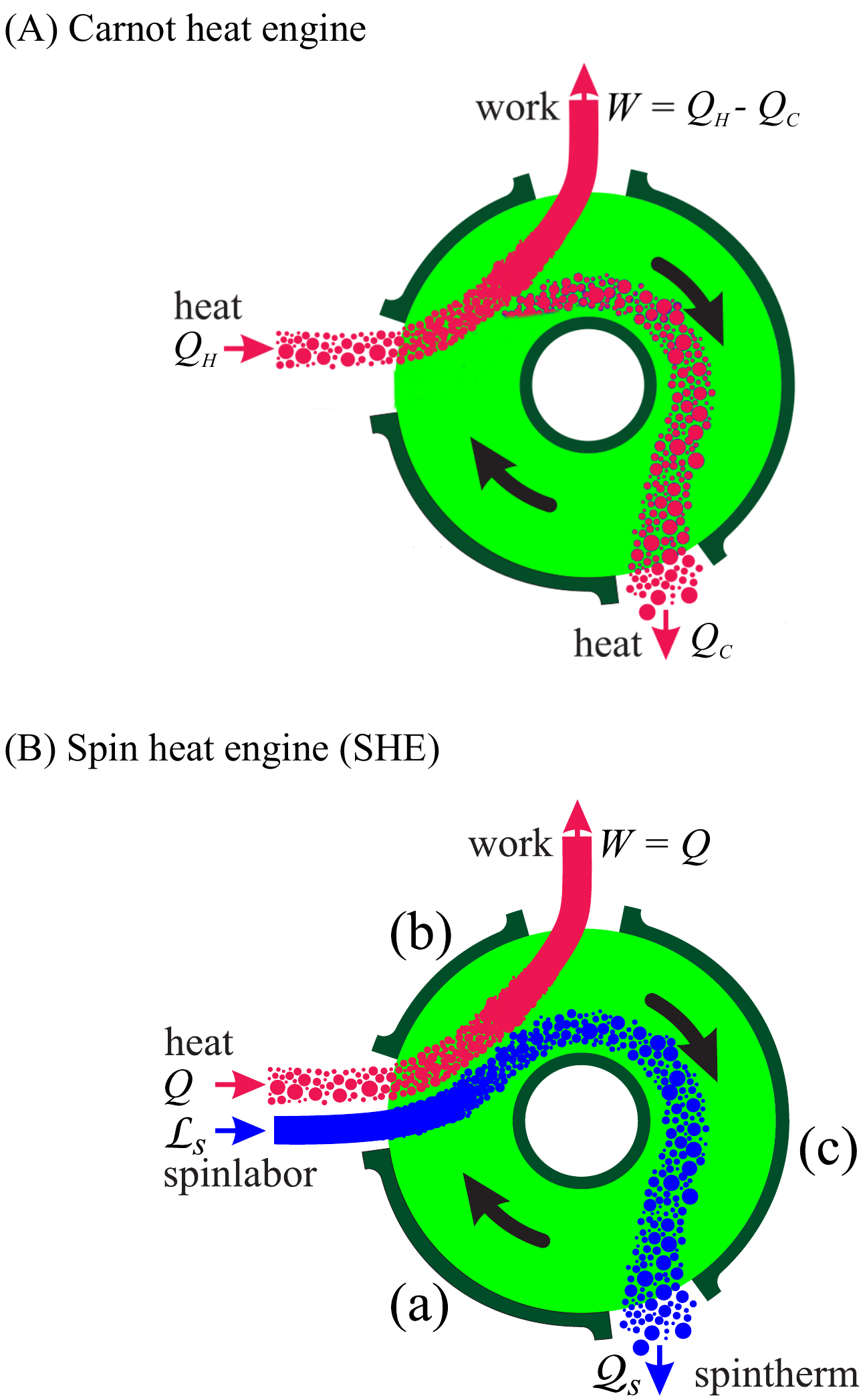}
    \caption{Conceptual diagrams for (A) a conventional Carnot
heat engine and (B) the VB spin-heat engine. In the Carnot
cycle, energy is transferred from a hot to a cold reservoir
producing useful work along the way. In the VB-SHE, work
is extracted from a single hot thermal reservoir. The cycle is
completed by resetting the system’s entropy in a process where
the initial spin supplied by an external source (the spinlabor
$\mathcal{L}_s$) is dissipated into a spin reservoir in the form of spintherm, $\mathcal{Q}_s$. The latter is the spin equivalent of the heat dissipated in the cold reservoir in the Carnot cycle.}
    \label{fig:SHE concept}
\end{figure}

A classical Carnot engine works between two thermal reservoirs at different temperatures as shown in the conceptual diagram of Fig. \ref{fig:SHE concept}-A. The working fluid absorbs heat ($Q_H$) from the hot reservoir, transforms part of this energy into work, and dumps the remaining energy into the cold reservoir ($Q_C$). The total work extracted is given by $W=Q_H-Q_C$ with an efficiency of $\epsilon=\frac{T_H-T_C}{T_H}$. 

The Vaccaro-Barnett SHE, on the other hand, uses a single thermal reservoir (see Fig.\ref{fig:SHE concept}-B). A full cycle is possible by inserting a spin-reservoir in place of the cold thermal reservoir. In this case, all the thermal energy coming from the hot reservoir ($Q$) is transformed into work ($W$). The remaining cost balance of the cycle does not involve energy, but rather another conserved quantity, the spin angular momentum. During the cycle, the spin equivalent of work (spinlabour $\mathcal{L}$) which is the change in angular momentum during the cycle, is dissipated as the spin equivalent of heat (spintherm $\mathcal{Q}$) into the spin reservoir. In the next section we show how this conceptual heat engine can be implemented using trapped ions.

\section{The ion-trap SHE}\label{sec: The ion-trap SHE}
The general scheme to implement an optical version of the SHE is depicted in Fig.\ref{fig: stages of SHE}. The working fluid is a three-level trapped ion with two energy-degenerate ground states that differ only by their spin. The ion is trapped in a harmonic potential and its vibrational degree of freedom is what mediates the heat exchange in the cycle. Initialized in the state $\rho=\ket{\uparrow}\bra{\uparrow}\otimes\rho_\nu (0)$, where $\rho_\nu(0)$ is a thermal vibrational state, the ion is subjected to a Raman pulse in the work extraction phase, as shown in Fig.\ref{fig: stages of SHE}-b. During this stage, energy from the motional states ($Q$) is transferred to the optical field in the form of useful work ($W$), at the same time that angular momentum from the initial spin distribution is used ($\mathcal{L}$) so that the transition to different spin states is possible. In the reset stage (Fig.\ref{fig: stages of SHE}-c), the system is brought into contact with a spin reservoir that removes entropy from the working fluid ($\mathcal{Q}$), bringing the electronic state back to $\ket{\uparrow}$. The system is then brought back into contact with the hot reservoir shown in Fig.\ref{fig: stages of SHE}-a so that the thermal motional state $\rho_\nu(0)$ is restored, closing the cycle. The resource balance during a full cycle obviously obey energy and angular momentum conservation laws such that $|Q| = |W |$ and $|\mathcal{L}| = |\mathcal{Q}|$. Furthermore, energy and angular momentum are interconnected by virtue of the work extraction stage represented in Fig.\ref{fig: stages of SHE}-b. The amount of work extracted will correspond to the vibrational energy converted in the process, which is given by the energy difference between the two Raman lasers $\delta$ multiplied by the probability of a successful transfer, i.e. the final population in level $\ket{\downarrow}$ ($P_\downarrow$):
\begin{equation}
    W=\hbar\delta P_\downarrow.\label{eq: Work as function of down population}
\end{equation}
Because, as explained above, for the transition to occur spin labor from the initial spin distribution needs to be dissipated as spintherm to account for the change in angular momentum. The spinlabour is the change in total angular momentum,
\begin{equation}
\mathcal{L}=\left<J_z(t)\right>-\left<J_z(0)\right>=\frac{\hbar}{2}(P_\uparrow-P_\downarrow)-\frac{\hbar}{2}=-\hbar P_\downarrow, \label{eq: Spin labor}
\end{equation}
where $J_z$ is the $z$ component of angular momentum. 
\section{Work extraction}\label{sec: Work extraction}
\subsection{The Model}\label{subsec: The model}
\begin{figure}
    \centering
    \includegraphics[width=\linewidth]{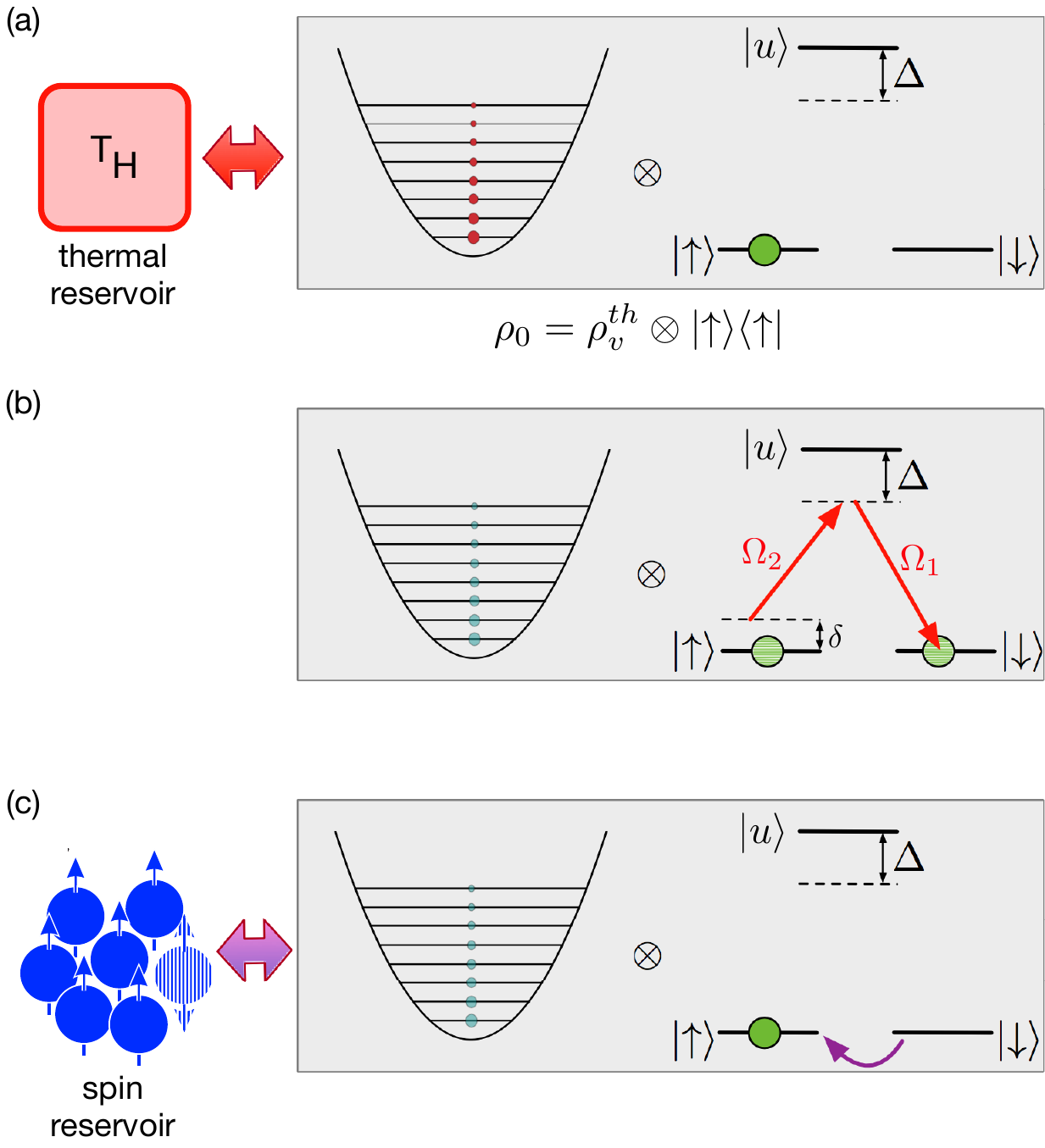}
    \caption{A schematic of an operation cycle of the optical SHE.
The working fluid comprises a trapped ion with three electronic levels and a quantized vibrational mode. The latter is
represented by the equally spaced levels of a harmonic oscillator, while the former contains two energy degenerate levels $\ket{\uparrow}$ and $\ket{\downarrow}$ and the upper state $\ket{u}$. Panel (a) represents the first stage of the engine cycle, where the working fluid is initialized in the spin up state (full green circle) and the vibrational state is in a thermal state due to the contact with the reservoir at temperature $T_H$. Panel (b) represents the work extraction stage, where energy from the thermal reservoir (coming from the vibrational degrees of freedom of the ion) is transferred in the form of coherent light (optical work). The resetting stage is represented in (c) where the system is brought into contact with a spin reservoir: the entropy in the internal spin states is then reset to zero at the expense of increasing the entropy in the spin reservoir.}\label{fig: stages of SHE}
\end{figure}

We will start the analysis of the work extraction stage by considering the Raman excitation scheme shown in Fig.\ref{fig: stages of SHE}-b. Levels $\ket{\uparrow}$ and $\ket{\downarrow}$ are energy-degenerate ground states corresponding to the different spins. The $\ket{\uparrow}\leftrightarrow\ket{u}$ and $\ket{\downarrow}\leftrightarrow\ket{u}$ transitions have frequency $\omega_0$ and the two lasers indicated in the figure are far detuned with frequencies $\omega_1=\omega_0-\Delta$ and $\omega_2=\omega_0-\Delta-\delta$, respectively. The Hamiltonian of the system is given by
\begin{equation}
\hat{H}=\hat{H}_0+\hat{H}_I,\label{eq: Full Raman hamiltonian}
\end{equation}
with
\begin{equation}
\hat{H}_0=\hbar\nu \hat{a}^\dagger \hat{a}+\hbar\omega_0\ket{u}\bra{u}, \label{eq: Harmonic hamiltonian}
\end{equation}
and
\begin{equation}
\begin{aligned}
    \hat{H}_I=&\hbar\Omega_2 e^{-i(\mathbf{k}_2\cdot \mathbf{x}-\omega_2t)}\ket{\uparrow}\bra{u}\\& + \hbar\Omega_1e^{-i(\mathbf{k}_1\cdot \mathbf{x}-\omega_1t)}\ket{\downarrow}\bra{u}+c.c. \label{eq: Interaction Hamiltonian}
\end{aligned}
\end{equation}
$H_I$ is the Hamiltonian describing the interaction of the ions with the two lasers in a traveling wave configuration. Here, $\mathbf{k}_i$ is the wave vector of the $i$\textsuperscript{th} laser and $\Omega_i$ is the corresponding Rabi frequency, assumed to be real and positive.

This three-level model can be simplified significantly if we consider the situation where the lasers are far detuned from level $\ket{u}$. In this situation, the transitions between levels $\ket{\uparrow}$ and $\ket{\downarrow}$ occur with negligible population of level $\ket{u}$, which can be removed from the model. More explicitly, when $\Delta\gg\delta,\Omega_i$, then we can adiabatically eliminate level $\ket{u}$ and obtain an effective two-level description with \cite{PhysRevA.53.2501}
\begin{equation}
\begin{aligned}
\hat{H}_\text{adia}&=\hbar\nu \hat{a}^\dagger \hat{a}-\hbar\frac{\Omega_1^2}{\Delta}\ket{\downarrow}\bra{\downarrow}-\hbar\frac{\Omega_2^2}{\Delta}\ket{\uparrow}\bra{\uparrow}-\hbar\delta\ket{\uparrow}\bra{\uparrow}\\
&-\hbar\frac{\Omega_1\Omega_2}{\Delta}\left(e^{-i\eta(\hat{a}+\hat{a}^\dagger)}\ket{\uparrow}\bra{\downarrow}+e^{i\eta(\hat{a}+\hat{a}^\dagger)}\ket{\downarrow}\bra{\uparrow}\right)\label{eq: Adiabatic Hamiltonian}
\end{aligned}
\end{equation}
where we have expressed $e^{i\delta\mathbf{k}\cdot\mathbf{x}}=e^{i\eta(\hat{a}+\hat{a}^\dagger)}$, where $\delta\mathbf{k}=\mathbf{k}_2-\mathbf{k}_2$, and $\eta=\hbar\sqrt{\delta k/2m}$ is the Lamb-Dicke parameter.  $\hat{a}^\dagger$, $\hat{a}$ are the creation and annihilation operators for the quantized motion of the ion respectively. 

Finally, we will consider that the system is in the resolved sideband regime, meaning that we can address individual vibrational levels through the laser interaction. If we choose $\delta'=\delta-\frac{\Omega_1^2}{\Delta}+\frac{\Omega_2^2}{\Delta}=\kappa \nu$, with $\kappa$ an integer, then the Raman coupling induces transitions between levels $\ket{\uparrow}$ and $\ket{\downarrow}$ that involve the exchange of exactly $\kappa$ vibrational quanta. In this case, we can write the final Hamiltonian representing the work extraction stage (in the interaction picture) as
\begin{equation}
\hat{H}=-\hbar\Omega(\hat{d}^\dagger\ket{\uparrow}\bra{\downarrow}+\hat{d}\ket{\downarrow}\bra{\uparrow}),\label{eq: Hamiltoniain after adiabatic limit and in interaction picture}
\end{equation}
with $\Omega=\Omega_1\Omega_2/\Delta$,
\begin{equation}
    \hat{d}=f_\kappa(\hat{a}^\dagger \hat{a})(i\eta \hat{a})^\kappa,
\end{equation}
and
\begin{equation}
    f_\kappa(\hat{a}^\dagger \hat{a})=e^{-\eta^2/2}\sum_{l=0}^{\infty}\frac{(-1)^l\eta^{2l}}{l!(\kappa+l)!}(\hat{a}^{\dagger}\hat{a})^l.
\end{equation}
Starting from Eq.\ref{eq: Hamiltoniain after adiabatic limit and in interaction picture}, and assuming that the ion is initially prepared in level $\ket{\uparrow}$ and with the vibrational degrees of freedom in a thermal state $\rho^{\text{th}}_\nu=\sum_mP(m)\ket{m}\bra{m}$, then we can write the solution for the expectation value of the number operator $\bar{n}(t)=\braket{\hat{a}^\dagger \hat{a}}$ explicitly (see Appendix \ref{sec: Analytic solution appendix}):
\begin{equation}
    \bar{n}(t)=\sum_m mP(m,t),\label{eq: Average motional state}
\end{equation}
where
\begin{equation}
\begin{aligned}
P(m,t)&=P(m)\cos^2\left(\Omega t\eta^\kappa\sqrt{\frac{m!}{(m-\kappa)!}}f_\kappa^{(m-\kappa)}\right)\\
&+P(m+\kappa)\sin^2\left(\Omega t\eta^\kappa\sqrt{\frac{(m+\kappa)!}{m!}}f_\kappa^{(m)}\right)\label{eq: Probability distribution of motional states}
\end{aligned}
\end{equation}
is the probability distribution of the number of excitations in the vibrational reservoir. Here, $P(m)\equiv P(m,0)=(1-e^{-\beta\hbar\nu})e^{-\beta m\hbar\nu}$ is the initial thermal distribution with $\beta=1/(k_BT_0)$. Alternatively, this distribution could be written in terms of the average number of vibrational quanta using the relation $\bar{n}_0=[e^{\beta\hbar\nu}-1]^{-1}$. 

\begin{figure}
    \centering
    \includegraphics[width=\linewidth]{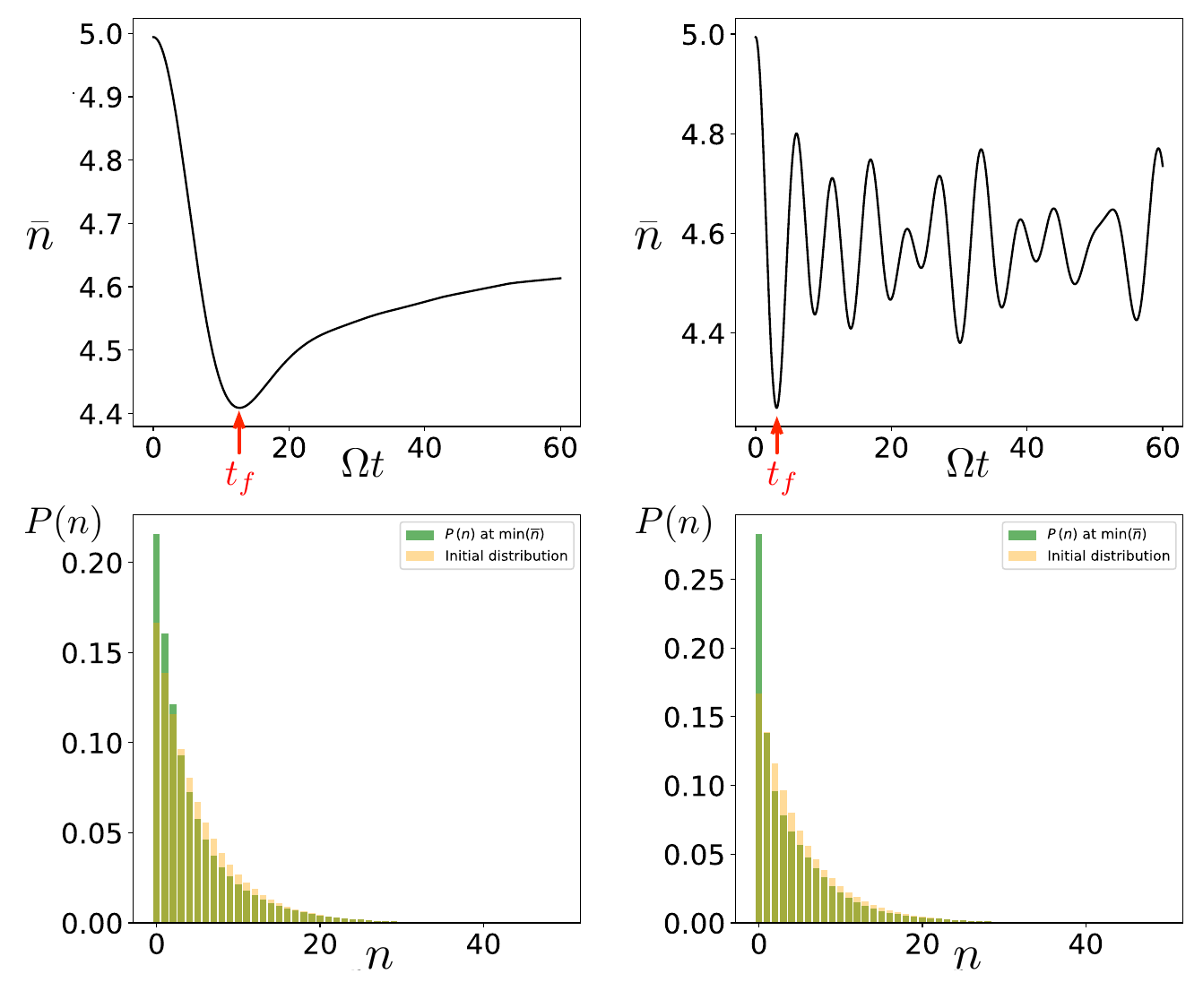}
    \caption{Top panels: Evolution of the average number of vibrational quanta as a function of time for $\eta=0.05$ (left) and 0.4 (right). The arrows represent the final time $t_f$ for the stage, chosen to be the one at which the work extracted is maximized ($\bar{n}$ is minimum). Bottom panels: Comparison between the probability distributions $P(n)$ at the beginning (orange) and at the end (green) of the stage}
    \label{fig: dynamics of nbar}
\end{figure}

The dynamics of this work extraction stage is shown in Fig.\ref{fig: dynamics of nbar} for $\eta=0.05$ (left panels) and $\eta=0.4$ (right panels). In both cases, we fixed the detuning by setting $\kappa=1$ and assumed a thermal reservoir with $\bar{n}_0=5$. Note that the work extracted, given by Eq.\eqref{eq: Work as function of down population}, can also be written in terms of the change in mean vibrational energy
\begin{equation}
    W=-\hbar\nu[\bar{n}-\bar{n}_0]\label{eq: Work as function of average quanta}
\end{equation}
and, therefore, can be obtained directly from the top panels in Fig.\ref{fig: dynamics of nbar}, where the time evolution of the mean number $\bar{n}(t)$ is shown. The initial thermal vibrational energy quickly decreases as it is transformed into optical work and we set the stage duration $(t_f)$ to be the first time at which $\bar{n}(t)$ reaches a minimum and, therefore, $W$ is maximum. During this process, the distribution $P(n)$ of the vibrational excitations cools down, remaining close to a thermal distribution at lower temperatures for small Lamb-Dicke parameters, but deviating considerably due to the nonlinearities present for larger $\eta$ (see bottom panels of Fig.\ref{fig: dynamics of nbar})

\subsection{Maximimizing work extraction}\label{subsec: Maximising work extraction}
To optimize the efficiency of a single extraction cycle, we should maximise $\Delta\bar{n}=\bar{n}_{t_f}-\bar{n}_0$. Three parameters are important here: the initial temperature of the reservoir, which indicates how much energy is available for conversion; the laser detuning, which dictates how many vibrational quanta will be involved in the transitions; and the Lamb-Dicke parameter, which controls the degree of nonlinearity of the function $f_\kappa^{(m)}$.

\begin{figure}
    \centering
    \includegraphics[width=\linewidth]{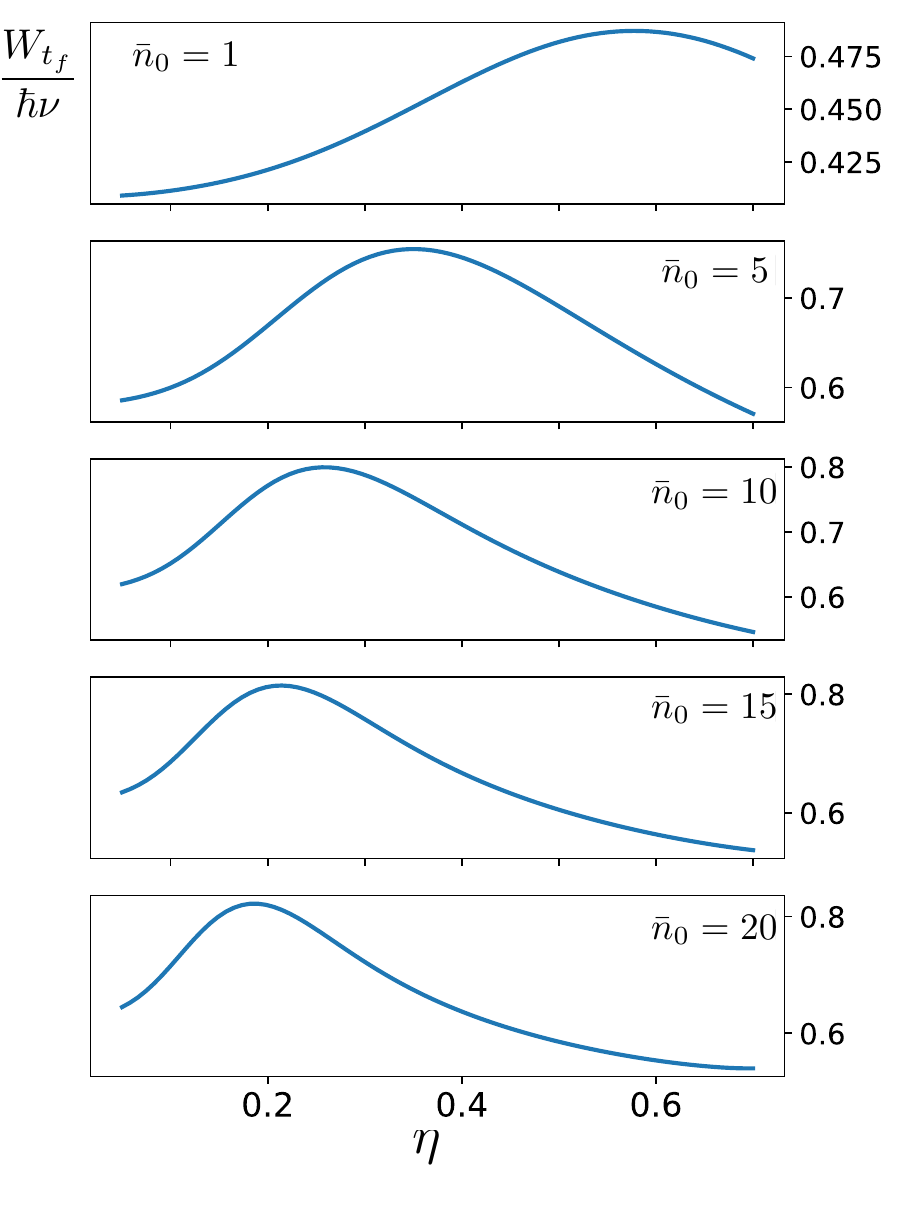}
    \caption{Work extracted in a cycle (in units of $\hbar\nu$) for $\kappa = 1$. The work ($W_{t_f}$) is calculated from the time evolution, as in Fig.\ref{fig: dynamics of nbar}), and plotted as a function of the Lamb-Dicke parameter for different values of the initial temperature. These curves show that there is an optimal $\eta$ ($\eta_\text{opt}$) that maximizes the work extracted.}
    \label{fig: Work extracted for different eta}
\end{figure}

\begin{figure}
    \centering
    \includegraphics[width=\linewidth]{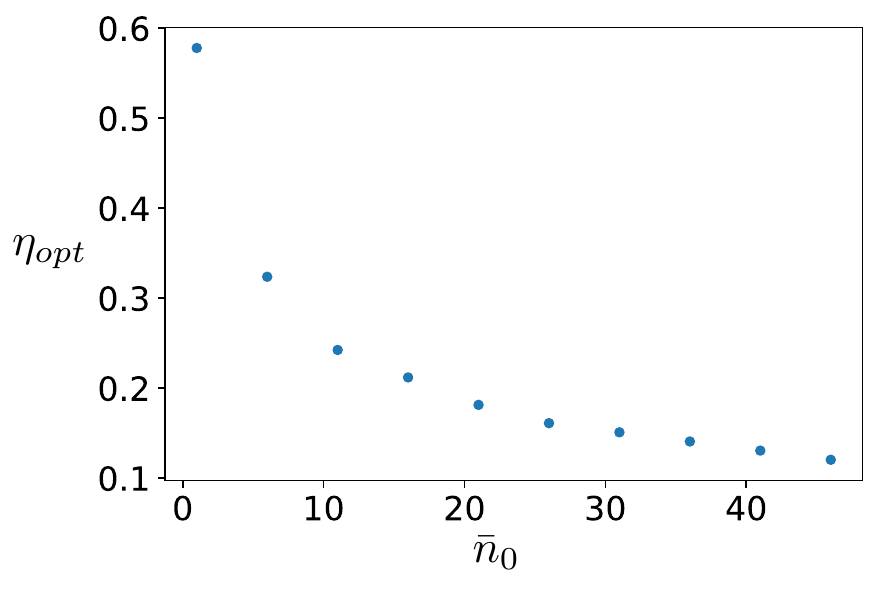}
    \caption{Optimal Lamb-Dicke parameter ($\eta_\text{opt}$, as defined in Fig. \ref{fig: Work extracted for different eta}) as a function of the initial average motional excitation $\bar{n}_0$. The curve shows that the higher the initial vibrational temperature, the smaller the value of $\eta_\text{opt}$.}
    \label{fig: optimal Lamb-dicke for nbar}
\end{figure}

We start our investigation by fixing the detuning by setting $\kappa=1$, as in Fig.\ref{fig: dynamics of nbar}, and analyzing the effects of $\eta$ and $\bar{n}_0$. In Fig.\ref{fig: Work extracted for different eta}, we plot the work extracted at the end of the cycle $(W_{t_f})$ as a function of $\eta$ for various reservoir temperatures. These results show that there is an optimum $\eta_{\text{opt}}$ that leads to $W_\text{opt}$ (given by the peaks in the curves) for each reservoir temperature, and that smaller $\eta$ is favored for larger $\bar{n}_0$. This is summarized in Fig.\ref{fig: optimal Lamb-dicke for nbar}, where we explicitly plot $\eta_\text{opt}$ as a function of $\bar{n}_0$.

Figure \ref{fig: Work extracted for different eta} also shows that $\Delta\bar{n}_{t_f}$ saturates as $\bar{n}_0$ increases. This should come as no surprise: even though this increase makes more energy available for conversion, the Raman transition tuned to the first sideband limits the amount of reservoir energy that can be harvested. The solution for that is to increase the laser detuning, allowing for more vibrational quanta to be involved in the Raman transitions. This is illustrated by the open symbols in Fig.\ref{fig: Optimal work extracted} where $W_\text{opt}$ is plotted against $\bar{n}_0$ for different values of $\kappa$. For small $\bar{n}_0$, transitions involving multiple quanta are suppressed, and tuning to lower sidebands is more efficient. However, as the initial temperature of the reservoir is increased, then larger detunings lead to higher values for the work extracted.

\subsection{Theoretical bounds on work extraction}\label{subsec: Theoretical bounds on work extraction}
Equation \ref{eq: Work as function of down population} shows that the maximum possible extracted work would correspond to the maximum value of $P_\downarrow=1$, which, by using Eq.\ref{eq: Work as function of average quanta}, leads to $\Delta\bar{n}_\text{max}=\kappa$. However, since we start with a thermal vibrational state, entropy will impose limits on what can be achieved. Considering that we have a bi-partite system (spin + vibrational), sub-additivity of entropy gives us
\begin{equation}
\Delta S_s+\Delta S_v \ge \Delta S_{sv}=0,\label{eq: Sub-adivity of entropy}
\end{equation}
\begin{figure}
    \centering
    \includegraphics[width=\linewidth]{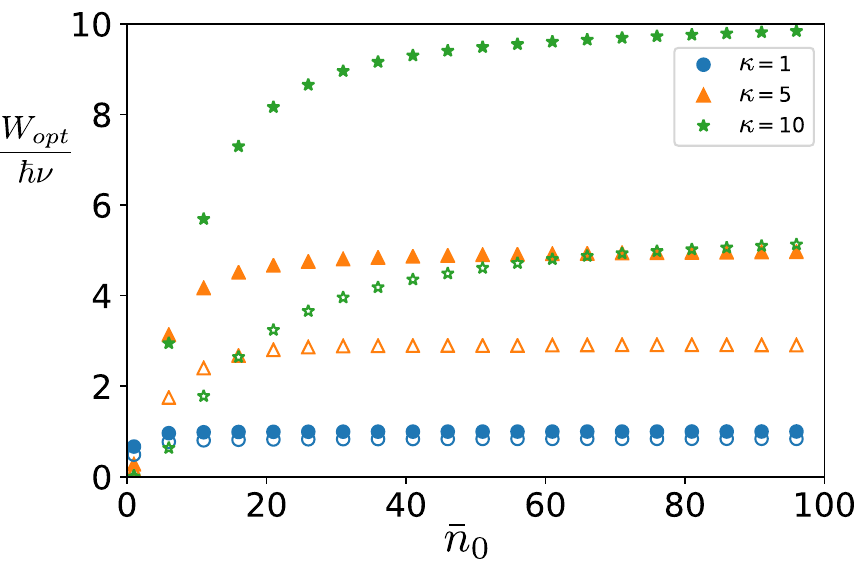}
    \caption{Optimal extracted work as a function of $\bar{n}_0$ for various values of the detuning parameter $\kappa$. Even though the initial available vibrational energy increases with $\bar{n}_0$, the useful work extracted is limited by the value of $\kappa$ and all curves eventually converge to their asymptotic value. For $\kappa=10$ one sees that the cycle is suppressed for low $\bar{n}_0$ as there is not enough initial energy to drive the transition. Curves are for $\kappa=1$ (blue), $\kappa=5$ (orange), and $\kappa=10$ (green). Full symbols correspond to the maximum work allowed by inequality \eqref{eq: Entropy inequality analytic} under the assumption of a final thermal state.}
    \label{fig: Optimal work extracted}
\end{figure}

\noindent where $S_i=-\text{Tr}[\rho_i\ln\rho_i]$ is the von Neumann entropy for the density matrix $\rho_i$. $\rho_{sv}$ is the total density matrix of the system, and $\rho_s=\text{Tr}_v[\rho_{sv}]$ and $\rho_v=\text{Tr}_s[\rho_{sv}]$ the reduced density matrices for the spin and vibrational degrees of freedom, respectively. The equality in the last equation comes from the fact that the total system is evolving under unitary dynamics.

For our initial state, $\rho_{sv}(0)=\ket{\uparrow}\bra{\uparrow}\otimes\rho_v^{\text{th}}(0)$, we have $S_s(0)=0$ and we can write
\begin{equation}
    -P_\downarrow\ln(P_\downarrow)-(1-P_\downarrow)\ln(1-P_\downarrow)+S_v(t_f)-S_v(\rho_v^\text{th}(0))\ge0.\label{eq: Entropy inequality analytic}
\end{equation}
Because $P_\downarrow$ and $\Delta\bar{n}$ are connected via Eqs.\eqref{eq: Work as function of down population} and \eqref{eq: Work as function of average quanta}, Eq.\eqref{eq: Entropy inequality analytic} then provides limits on $\bar{n}_f$ or, equivalently, the possible values of $P_\downarrow$, given an initial mean vibrational quanta $\bar{n}_0$. 

\begin{figure}
    \centering
    \includegraphics[width=\linewidth]{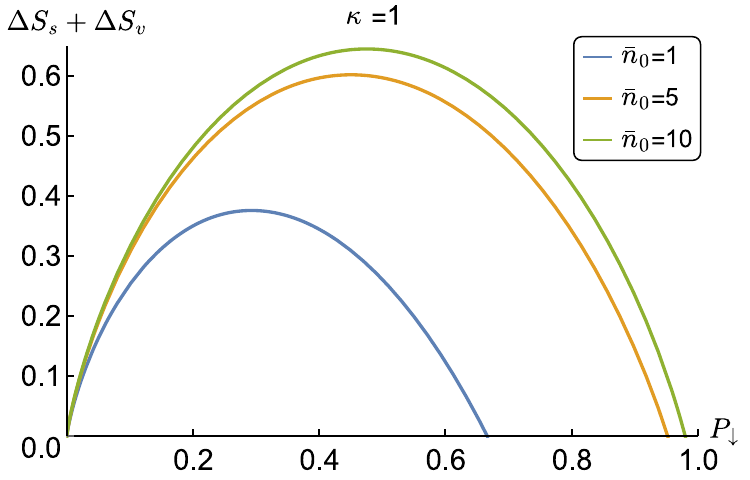}
    \caption{Sum of the von Neumann entropies for the reduced density matrices as a function of $P_\downarrow$ for different values of initial vibrational quanta $\bar{n}_0$ and $\kappa=1$}
    \label{fig: Von neuman entropies example}
\end{figure}

To have an idea of this bound, we consider the situation where the final vibrational state is also a thermal state. As we have already seen in Fig.\ref{fig: dynamics of nbar}, this is not true in general, but it is a good approximation in some regimes and allows us to analytically calculate the entropies in Eq.\eqref{eq: Entropy inequality analytic}. Figure \ref{fig: Von neuman entropies example} shows the left-hand-side of Eq.\eqref{eq: Entropy inequality analytic}.

By calculating the roots of these curves, we can calculate the maximum value of $P_\downarrow$, thus the maximum work, allowed by the entropy inequality, Eq.\eqref{eq: Entropy inequality analytic}. This is shown by the full symbols in Fig.\ref{fig: Optimal work extracted} and should be compared with the results obtained from our full dynamical model (open symbols).

\section{Spin-reset stage: spin-polarized reservoir}\label{sec: Spin-Reset stage: Spin polarised reservoir}
At the end of the work-extraction stage, the spin and
vibrational degrees of freedom of the ion are entangled
and the reduced density matrix for the spin variables is
left in a diagonal mixed state with the population spread
between the $\ket{\uparrow}$ and $\ket{\downarrow}$ states. This was already alluded to in Fig.\ref{fig: stages of SHE}-b, and is also evident from Eq.\ref{eq: Pup state over time} in Appendix \ref{subsec: Density matrix for the electronic degree of freedom}. At this point, to be able to extract more work, one needs to reset the system to its initial state, closing the cycle. Our strategy to accomplish this reset, or erasure, step is a passive one: we consider that the ion is coupled to a spin reservoir that, through spin-exchange collisions, transfers electrons from $\ket{\downarrow}$ to $\ket{\uparrow}$, as illustrated in Fig.\ref{fig: stages of SHE}-c. The reservoir acts as an entropy sink, removing entropy from the system as it brings it back to the pure state $\ket{\uparrow}$. Mathematically, this perfectly polarized spin bath gives a resetting dynamics that can be described by a master equation of the form
\begin{equation}
    \dot{\rho}=\Gamma_s\mathcal{D}[\ket{\uparrow}\bra{\downarrow}]\rho,\label{eq: Spin reset linblad equation}
\end{equation}
where $\Gamma_s$ is the spin decay rate, $\mathcal{D}[L]\rho=L\rho L^\dagger-1/2(L^\dagger L\rho+\rho L^\dagger L)$ is the decoherence superoperator in the Lindblad form \cite{lindblad1976generators}. In our case, the spin populations would start with their values at the end of the work extraction stage ($t=t_f$) and the solution $P_\downarrow=e^{-\Gamma_s t}P_\downarrow(t_f)$, corresponding to a simple exponential decay, would lead to the electronic population being transferred back to state $\ket{\uparrow}$ in the asymptotic limit. 

This scenario, seemingly idealized, has already been
implemented experimentally in a different context. In
their experiments, Ratschbacher et al. \cite{ratschbacher2013decoherence} surrounded a single trapped Yb$^+$ ion by a cloud of spin-polarized
ultracold $^{87}$Rb neutral atoms and observe the effect of
atom-ion collisions in the spin dynamics. For $^{171}$Yb$^+$
interacting with Rb atoms initially prepared in the hyperfine $\ket{F=1, m_F=-1}$ state, the ions relax to the spin
ground state in a time scale given by $T_1 \approx 8.7$ ms. Note,
however, that their result does not directly map to our
SHE scheme. For example, the $^{171}$Yb$^+$ levels used in \cite{ratschbacher2013decoherence} are not degenerated, but rather the $\ket{F=1, m_F=0}$
and $\ket{F=0, m_F=0}$ states. To really demonstrate the
SHE with no energy cost, one would need to choose a
different pair of states and check whether the resulting
spin-exchange collisions still work as well as shown in \cite{ratschbacher2013decoherence}. Also, in their experiment the temperature of the ion is limited by micromotion heating, which is of the order of 20 mK. This would correspond to an average number of thermal quanta of $\bar{n}\approx2700$, which is much larger than what we discussed in Sec. \ref{sec: Work extraction}. This doesn’t seem to be a major problem, as our calculations showed that increasing $\bar{n}_0$ seem to be better for work extraction (at least for moderate $\bar{n}_0$).

\section{Re-thermalization stage}\label{sec: Re-thermalisation stage}
After extracting energy from the vibrational mode and
re-setting the spin, the final step is to bring the ion back
in contact with the hot reservoir to re-thermalize and return to its initial thermal state. This stage, shown in Fig.\ref{fig: stages of SHE}-a, will be modeled using a thermal master equation for the motional states:
\begin{equation}
    \dot{\rho}_v=\Gamma_H(\bar{n}_0+1)\mathcal{D}[\hat{a}]\rho_v + \Gamma_H\bar{n}_0\mathcal{D}[\hat{a}^\dagger]\rho_v.\label{eq: Re-thermalisation linblad equation}
\end{equation}
Under this equation, the system asymptotically reaches
the initial thermal state $\rho_v^{\text{th}}$.

\section{Full engine cycle}\label{sec: Full engine cycle}
\begin{figure}[t]
    \centering
    \includegraphics[width=\linewidth]{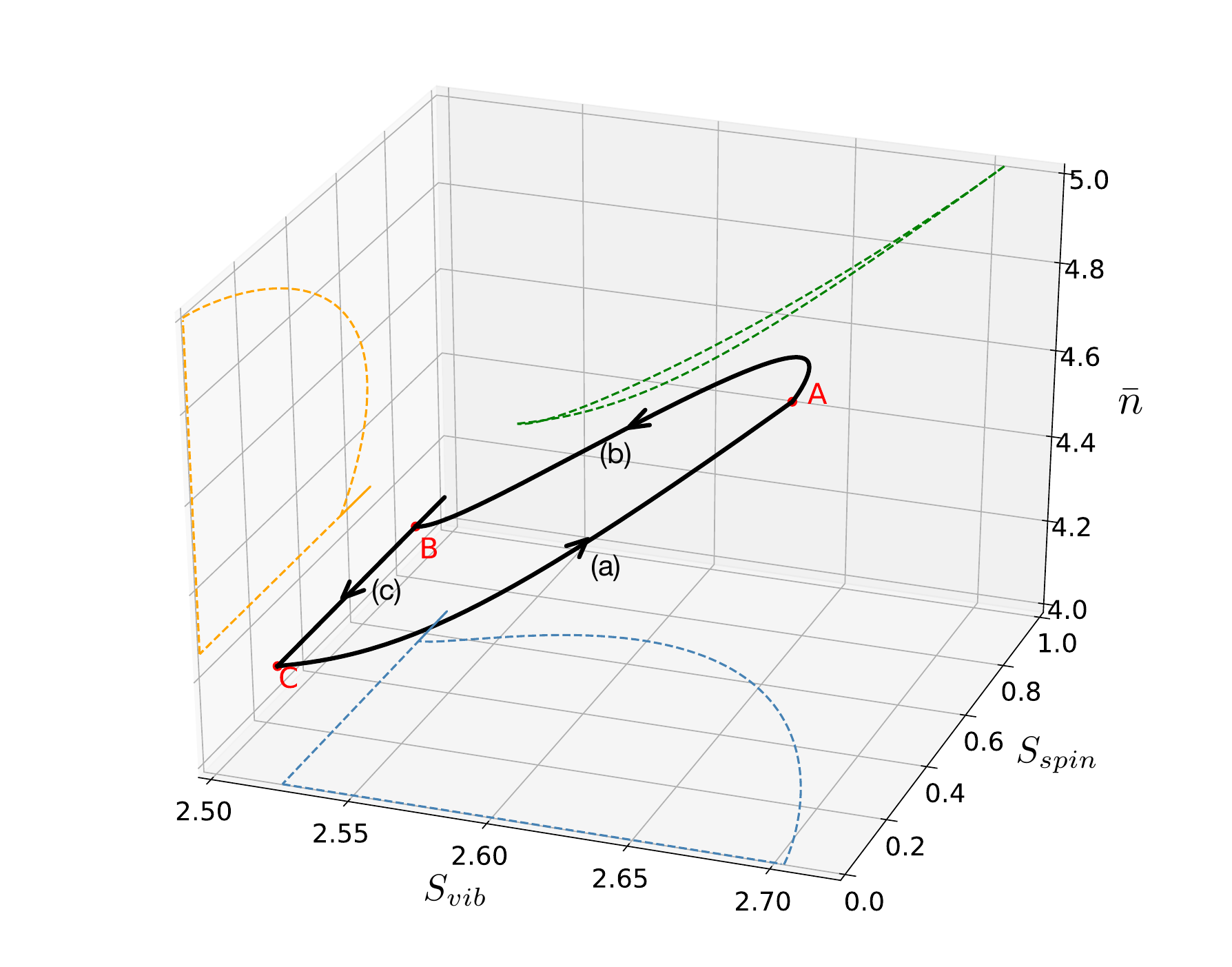}
    \caption{SHE cycle represented in a three-dimensional T-S diagram. The vertical axis correspond to $\bar{n}$ (parameterizing the temperature) while the other axes represent entropies for the spin ($S_\text{spin}$) and for the vibrational ($S_\text{vib}$) degrees of freedom. Work extraction, spin-reset, and re-thermalisation stages represent, respectively, the processes (b), (c), and (a) in Fig. \ref{fig: stages of SHE}}
    \label{fig: full engine cycle}
\end{figure}

A simulation of a full cycle of the SHE is shown in
Fig.\ref{fig: full engine cycle}, where we have used $\bar{n}_0$, $\eta=0.4$, and $\kappa=1$.
The system starts at point A with the ion in the initial
state $\rho=\ket{\uparrow}\bra{\uparrow}\otimes\rho_v(0)$. It then undergoes the work extraction phase (b) described in Section \ref{sec: Work extraction}, with the curve from $A$ to $B$ given by our analytical solution in Eqs. \eqref{eq: Average motional state}, \eqref{eq: Probability distribution of motional states}, (\eqref{eq: Pup state over time}, \eqref{eq: Entropy analytic appendix}). From $B$ to $C$, the system undergoes the spin-reset stage described by Eq. \eqref{eq: Spin reset linblad equation}, where we let the system evolve long enough ($t=10/\Gamma_s$) so that it is close to its asymptotic spin state. Note that since the population in state $\ket{\downarrow}$ at $B$ is larger than $1/2$, the system initially increases its spin entropy before reducing it back to zero when it reaches the $\ket{\uparrow}$ state. Finally, from $C$ to $A$, the system evolves under the re-thermalisation process given by Eq. \eqref{eq: Re-thermalisation linblad equation} with $\bar{n}_0=5$ and $t = 20/\Gamma_H$. Since this step is a purely thermal process, the spin entropy remains unchanged, but the vibrational entropy and temperature return to their value at the beginning of the cycle. The projections in the different planes, indicated by the dashed lines, are included to help visualizing the process. Note that the projections on the $S_v-\bar{n}$ do not coincide, indicating the differences between the
$A \rightarrow B$ and $C \rightarrow A$ processes: the latter is pure thermalisation with the hot reservoir while the former is a
non-equilibrium process induced by the Raman transition.

\section{The exchange of angular momentum- and energy-type works}
The free entropy concept by Guryanova et al. \cite{Guryanova_2016} is the entropic equivalent to free energy for thermodynamic systems with multiple conserved quantities. As we will soon prove, the QSHE obeys the change in free entropy inequality
\begin{equation}
\lambda_{s}\tilde{W}_s+\lambda_{v}\tilde{W}_{v}\le-\Delta \tilde{F}\label{eq: Free entropy},
\end{equation}
where $\lambda_i$ and $\tilde{W}_i$ are the unitless inverse temperature and unitless work for bath $i$, with the latter defined as the change in unitless conserved quantity $A_i$, i.e. $\tilde{W}_i=-\Delta A_i$, and $\Delta\tilde{F}=\tilde{F}(\rho)-\tilde{F}(\rho_0)$ is the change in the system's free entropy. Here the subscripts $s$ and $v$ represent the angular momentum and energy conserved quantities. The free entropy itself is defined by
\begin{equation}
    \tilde{F}=\sum_{i}\beta_i\braket{A_i}-S(\rho).
\end{equation}
The system in the QSHE is the ion, while the baths are the thermal and spin reservoirs used for resetting the ion to it's initial state. As $\Delta S=S(\rho)-S(\rho_0)=0$ under unitary evolution, then eq.\eqref{eq: Free entropy} saturates the inequality.

The initial state used throughout this manuscript $\rho_0=\ket{\uparrow}\bra{\uparrow}\otimes\rho_{th}$ implies $\lambda_{s}=\infty$. After the spin-reset and re-thermalization stages of the engine cycle, the ion returns to it's initial state, and the free entropy $\Delta\tilde{F}=0$. For Eq.\eqref{eq: Free entropy} to remain true with a finite $\lambda_{v}$, then $\tilde W_{v}=\infty$, implying that the QSHE must extract an infinite amount of work from the thermal energy reservoir. This result is based on the assumption that the ion is reset to it's original state after spin-reset and re-thermalisation, however for finite baths, this reset will instead be imperfect, and the assumption of constant $\lambda$'s no longer applies.

Eq.\eqref{eq: Free entropy} is general, and applies to any combination of conserved quantities whether they commute or not \cite{Guryanova_2016}. The $\infty$ work required to fulfill the inequality assuming the system is reset to it's initial, completely spin polarized state determines not only the amount of work that the QSHE can extract, but also the amount of work needed to initially prepare the system. Therefore, it takes infinite resources to prepare the perfectly polarized system, where these resources are arbitrary conserved quantities in \cite{Guryanova_2016}. This is commonly experienced in experiment, where 100\% purity during state preparation has never been achieved due to finite resource availability (limited time, energy, etc.). 

Eq.\eqref{eq: Free entropy} is more insightful when the inverse spin temperature is finite. Applying Jaynes' maximum entropy principle to the spin states, the initial spin distribution becomes thermal:
\begin{equation}
\rho_s=\frac{1}{1+e^{-\lambda_s}}\ket{\uparrow}\bra{\uparrow}+\frac{e^{-\lambda_s}}{1+e^{-\lambda_s}}\ket{\downarrow}\bra{\downarrow}.
\end{equation}
The unitless spin labor $\tilde{W}_{s}=\mathcal{L}/\hbar$ from Eq.\eqref{eq: Spin labor} is now $P_\downarrow(0)-P_\downarrow(t)$, while the unitless work $\tilde{W}_{v}=W/\kappa h\nu$ continues to be defined by Eq.\eqref{eq: Work as function of average quanta}. We show analytically in Appendix \ref{subsec: Equality of A and B-type work in the QSHE} that $\tilde{W}_s=\tilde{W}_{v}$ for any initial state. What is particularly informative is that for an initially `hot' thermal energy reservoir, as compared to the unitless spin temperature $\lambda_{v}^{-1}=\tau_{v}>\tau_s=\lambda_s^{-1}$, then $\tilde{W}_{v}, \tilde{W}_s\le0$. After spin-reset and re-thermalization with the baths, $\Delta \tilde{F}=0$. The changes in work are now entirely representative of the changes in angular momentum and energy in the baths. The energy of the thermal energy bath decreases, and it's average energy tends closer to 0 than it did prior to the QSHE cycle. Interpreting the distribution of energy in terms of it's polarization, then the thermal energy bath becomes more polarized. Meanwhile spin reservoir depolarizes. The QSHE is therefore an explicit example of work exchange between different conserved quantities.

\section{Discussion and Conclusion}
The implementation of the ion trap spin heat engine proposed here is an important first step to demonstrate the principles of heat engines that work beyond the usual paradigm of machines that require two thermal energy reservoirs to run. From the perspective of recent work \cite{mcclelland2025beyond}, it provides the working mechanism that would allow for beyond-Carnot energy efficiencies in ensembles with multiple conserved quantities, and realizes the trade-offs of quantum coherence between arbitrary conserved quantities \cite{Guryanova_2016}.

\section{Acknowledgements}
This research was supported by the ARC Linkage
Grants No. LP140100797, LP180100096 and the Lockheed Martin Corporation. We would like to acknowledge helpful discussions with N. Allen, L. Uribarri, A. Jacombhood, R. Glen, E. Cavalcanti and T. Gould. 

\appendix
\section{Analytic solution}
\label{sec: Analytic solution appendix}
To be able to address higher temperatures, we analyze the solutions for the dynamics under the Hamiltonian given in Eq.\eqref{eq: Hamiltoniain after adiabatic limit and in interaction picture}:
\begin{equation}
    \hat{H}=\hbar\Omega(\hat{d}^\dagger \ket{\uparrow}\bra{\downarrow}+\hat{d}\ket{\downarrow}\bra{\uparrow}).
\end{equation}

Expanding the evolution operator $\hat{U}=e^{-i\hat{H}t/\hbar}$ in a power series
\begin{equation}
    \hat{U}=1+i\Omega t\mathcal{H} + \frac{(i\Omega t)^2}{2!}\mathcal{H}^2 + \frac{(i\Omega t)^3}{3!}\mathcal{H}^3+\dots,
\end{equation}
where we have defined $\mathcal{H}\equiv \hat{d}^\dagger\ket{\uparrow}\bra{\downarrow}+\hat{d}\ket{\downarrow}\bra{\uparrow}$, and  writing the powers of this operator in a simpler form
\begin{align}
    i)\mathcal{H} &= \hat{d}^\dagger\ket{\uparrow}\bra{\downarrow}+\hat{d}\ket{\downarrow}\bra{\uparrow}\\
    ii) \mathcal{H}^2 &=\hat{d}^\dagger \hat{d}\ket{\uparrow}\bra{\uparrow}+\hat{d}\hat{d}^\dagger\ket{\downarrow}\bra{\downarrow}\\
    iii)\mathcal{H}^3 &=\hat{d}^\dagger \hat{d}\hat{d}^\dagger\ket{\uparrow}\bra{\downarrow}+\hat{d}\hat{d}^\dagger \hat{d}\ket{\downarrow}\bra{\uparrow}\\
    iv) \mathcal{H}^4 &=\hat{d}^\dagger \hat{d}\hat{d}^\dagger \hat{d}\ket{\uparrow}\bra{\uparrow}+\hat{d}\hat{d}^\dagger \hat{d}\hat{d}^\dagger\ket{\downarrow}\bra{\downarrow},\\
    &\vdots\notag
\end{align}
we can then write the evolution operator as
\begin{equation}
    \begin{aligned}
    \hat{U}&=\ket{\uparrow}\bra{\uparrow}\left(1+\frac{(i\Omega t)^2}{2!}\hat{d}^\dagger \hat{d}+\frac{(i\Omega t)^4}{4!}(\hat{d}^\dagger \hat{d})^2+\dots\right)\\
    &+\ket{\downarrow}\bra{\downarrow}\left(1+\frac{(i\Omega t)^2}{2!}\hat{d}\hat{d}^\dagger+\frac{(i\Omega t)^4}{4!}(\hat{d}\hat{d}^\dagger )^2+\dots\right)\\
    &+\ket{\uparrow}\bra{\downarrow}\left(i\Omega t \hat{d}^\dagger + \frac{(i\Omega t)^3}{3!}\hat{d}^\dagger \hat{d}\hat{d}^\dagger+\dots\right)\\
    &+\ket{\downarrow}\bra{\uparrow}\left(i\Omega t \hat{d} + \frac{(i\Omega t)^3}{3!}\hat{d}\hat{d}^\dagger \hat{d}+\dots\right).
    \end{aligned}
\end{equation}
This simplifies to
\begin{equation}
\begin{aligned}
    \hat{U}&=\ket{\uparrow}\bra{\uparrow}\cos\left(\Omega t\sqrt{\hat{d}^\dagger \hat{d}}\right) + \ket{\downarrow}\bra{\downarrow}\cos\left(\Omega t\sqrt{\hat{d} \hat{d}^\dagger}\right)\\
    &+i\ket{\uparrow}\bra{\downarrow}\hat{d}^\dagger\frac{\sin\left(\Omega t\sqrt{\hat{d}\hat{d}^\dagger}\right)}{\sqrt{\hat{d} \hat{d}^\dagger}} + i\ket{\downarrow}\bra{\uparrow}\hat{d}\frac{\sin\left(\Omega t\sqrt{\hat{d}^\dagger \hat{d}}\right)}{\sqrt{\hat{d}^\dagger \hat{d}}}.\\\label{eq: Full simplified unitary operator}
    \end{aligned}
\end{equation}
Assuming that we start with the ion in the electronic
state $\ket{\uparrow}$ and a thermal vibrational state, such that $\rho_0=\ket{\uparrow}\bra{\uparrow}\otimes\rho_\text{th}$, then the state at time $t$, $\rho(t)=\hat{U}\rho_0 \hat{U}^\dagger$, will be:
\begin{equation}
\begin{aligned}
    \rho(t)=&\left[\ket{\uparrow}\cos\left(\Omega t\sqrt{\hat{d}^\dagger \hat{d}}\right) + i\ket{\downarrow} \hat{d}\frac{\sin\left(\Omega t\sqrt{\hat{d}^\dagger \hat{d}}\right)}{\sqrt{\hat{d}^\dagger \hat{d}}}\right]\rho_\text{th}\\
    &\times\left[\bra{\uparrow}\cos\left(\Omega t \sqrt{\hat{d}^\dagger \hat{d}}\right)-i\bra{\downarrow}\frac{\sin\left(\Omega t\sqrt{\hat{d}^\dagger \hat{d}}\right)}{\sqrt{\hat{d}^\dagger \hat{d}}}\hat{d}^\dagger\right].\label{eq: Full density matrix evolution}
\end{aligned}
\end{equation}
\newpage
\subsection{Average vibrational excitation}\label{sec: Average vibrational excitation appendix}
The average vibrational excitation at time $t$ will be
\begin{widetext}
\begin{equation}
\begin{aligned}
    &\bar{n}(t) = \text{Tr}[\hat{n}\rho(t)]\\
    =&\sum_m m \left(\bra{m}\cos\left(\Omega t\sqrt{\hat{d}^\dagger \hat{d}}\right)\rho_\text{th}\cos\left(\Omega t\sqrt{\hat{d}^\dagger \hat{d}}\right)\ket{m}    +\bra{m}\hat{d}\frac{\sin\left(\Omega t \sqrt{\hat{d}^\dagger \hat{d}}\right)}{\sqrt{\hat{d}^\dagger \hat{d}}}\rho_\text{th}\frac{\sin\left(\Omega t\sqrt{\hat{d}^\dagger \hat{d}}\right)}{\sqrt{\hat{d}^\dagger \hat{d}}}\hat{d}^\dagger\ket{m}\right)\\
    =&\sum_{m,n}mP(n)\left[\bra{m}\cos\left(\Omega t \sqrt{\hat{d}^\dagger \hat{d}}\right)\ket{n}\bra{n}\cos\left(\Omega \sqrt{\hat{d}^\dagger \hat{d}}\right)\ket{m}+\bra{m}\hat{d}\frac{\sin\left(\Omega t\sqrt{\hat{d}^\dagger \hat{d}}\right)}{\sqrt{\hat{d}^\dagger \hat{d}}}\ket{n}\bra{n}\frac{\sin\left(\Omega t\sqrt{\hat{d}^\dagger \hat{d}}\right)}{\sqrt{\hat{d}^\dagger \hat{d}}}\hat{d}^\dagger\ket{m}\right]
\end{aligned}
\end{equation}
\end{widetext}

We can now use the fact that $f_\kappa$ and $d^\dagger d$ are diagonal in the number basis to simplify it further

$$
\begin{aligned}
    \bar{n}(t)
    &=\sum_mmP(m)\bra{m}\cos^2\left(\Omega t\sqrt{\hat{d}^\dagger \hat{d}}\right)\ket{m}
    \\&+\sum_{m}(m-\kappa)P(m)\eta^{2\kappa}\frac{m!}{(m-\kappa)!}\left(f_\kappa^{(m-\kappa, m-\kappa)}\right)^2\\&\times\bra{m}\frac{\sin^2\left(\Omega t\sqrt{\hat{d}^\dagger \hat{d}}\right)}{\hat{d}^\dagger \hat{d}}\ket{m}.
\end{aligned}
$$
What is needed now is to calculate the matrix elements
of $f_\kappa$ and $\hat{d}^\dagger \hat{d}$. They can be written explicitly as
\begin{equation}
\begin{aligned}
    \bra{m}\hat{d}^\dagger &\hat{d}\ket{n}=\eta^{2\kappa}\bra{m}(\hat{a}^\dagger)^\kappa\left(f_\kappa(\hat{a}^\dagger \hat{a})\right)^2\ket{n}\\
    &=\eta^{2\kappa}\sqrt{\frac{m!}{(m-\kappa)!}}\sqrt{\frac{n!}{(n-\kappa)!}}\left(f_\kappa^{(m-\kappa.n-\kappa)}\right)^2\delta_{m,n}\\
    &=\begin{cases}
        \delta_{m,n}\eta^{2\kappa}\frac{m!}{(m-\kappa)!}\left(f_\kappa^{(m-\kappa,n-\kappa)}\right)^2, & \text{for }m\ge\kappa\\
        0, &\text{for }m<\kappa\label{eq: Matrix elements of d^dagger d}
    \end{cases}
\end{aligned}
\end{equation}
and
\begin{equation}
\begin{aligned}
    f_\kappa^{(m,n)}&=\bra{m}f_\kappa\ket{n}=\bra{m}e^{-\eta^2/2}\sum_{l=0}^\infty \frac{(-1)^l\eta^{2l}}{l!(\kappa+l)!}(\hat{a}^\dagger \hat{a})^l\ket{n}\\
    &=\delta_{m,n}e^{-\eta^2/2}\sum_{l=0}^n\frac{(-1)^l\eta^{2l}}{l!(\kappa+l)!}\frac{n!}{(n-l)!}\\
    &=\delta_{m,n}e^{-\eta^2/2}\frac{n!}{(n+\kappa)!}L_n^\kappa(\eta^2)
\end{aligned}
\end{equation}
where $L_n^\kappa(x)$ are the generalized Laguerre polynomials
\begin{equation}
L_n^\kappa(x)=\sum_{l=0}^n\frac{(-x)^l}{l!}\frac{(n+\kappa)!}{(\kappa+l)!(n-l)!}.
\end{equation}
Note that the result in Eq.\eqref{eq: Matrix elements of d^dagger d} appears in the arguments of the sine and cosine terms in Eq.\eqref{eq: Full density matrix evolution} and therefore they are properly defined for all values of $m$.

Replacing \eqref{eq: Matrix elements of d^dagger d} in the expression for $\bar{n}(t)$ we have
\begin{equation}
    \begin{aligned}
        \bar{n}(t)&=\sum_{m}P(m)\left[m\cos^2\left(\Omega t\eta^\kappa\sqrt{\frac{m!}{(m-\kappa)!}}f_\kappa^{(m-\kappa)}\right)\right.\\
        &\left.+(m-\kappa)\sin^2\left(\Omega t\eta^\kappa\sqrt{\frac{m!}{(m-\kappa)!}}f_\kappa^{(m-\kappa)}\right)\right].
    \end{aligned}
\end{equation}
Alternatively, we can write it in terms of the probability
distribution for the vibrational energy at time t. Using that $\bar{n}(t)=\sum_mmP(m,t)$, we have
\begin{equation}
    \begin{aligned}
        P(m,t)&=P(m)\cos^2\left(\Omega t\eta^\kappa\sqrt{\frac{m!}{(m-\kappa)!}}f_\kappa^{(m-\kappa)}\right)\\
        &+P(m+\kappa)\sin^2\left(\Omega t\eta^\kappa\sqrt{\frac{(m+\kappa)!}{m!}}f_\kappa^{(m)}\right).
    \end{aligned}
\end{equation}

\subsection{Density matrix for the electronic degree of freedom}\label{subsec: Density matrix for the electronic degree of freedom}
For the initial conditions considered, the off-diagonal elements will be zero (unless $\kappa=0$, in which case there is no heat extraction) and we only need to consider the diagonal terms. The populations in levels $\ket{\uparrow}$ ($P_\uparrow$) and $\ket{\downarrow}$ ($P_\downarrow$=$1-P_\uparrow$) can be obtained straightforwardly from eq.\eqref{eq: Full density matrix evolution}:
\begin{align}  P_\uparrow&=\text{Tr}\left[\ket{\uparrow}\bra{\uparrow}\rho(t)\right]=\text{Tr}\left[\cos^2\left(\Omega t\sqrt{\hat{d}^\dagger \hat{d}}\right)\rho_{\text{th}}\right]\notag\\
    &=\sum_m P(m)\bra{m}\cos^2\left(\Omega t\sqrt{\hat{d}^\dagger \hat{d}}\right)\ket{m}\notag\\
    &=\sum_{m=0}^\infty P(m)\cos^2\left(\Omega t\eta^\kappa\sqrt{\frac{m!}{(m-\kappa)!}}f_\kappa^{(m-\kappa)}\right)\label{eq: Pup state over time}
\end{align}
where, in the last line, we used the result from Eq. \eqref{eq: Matrix elements of d^dagger d}. From the populations we can immediately extract the von Neumann entropy:
\begin{equation}
    S=-\rho\ln\rho = -P_\uparrow\ln P_\uparrow - P_\downarrow\ln P_\downarrow.\label{eq: Entropy analytic appendix}
\end{equation}

\subsection{Equality of A and B-type work in the QSHE}\label{subsec: Equality of A and B-type work in the QSHE}
For non-infinite inverse spin temperature, the spin and motional sub-systems evolve according to
\begin{equation}
\begin{aligned}
    P_\uparrow(t)&=\sum_{m<\kappa} p_\uparrow P(m)+\sum_{m\ge\kappa}p_\uparrow P(m)\cos^2\left(\Omega_m t\right)\\&+ \sum_m p_\downarrow P(m)\sin^2\left(\Omega_{m+\kappa} t \right)
\end{aligned}
\end{equation}
\begin{equation}
\begin{aligned}
P(m,t)&=p_\uparrow\left[P(m)\cos^2\left(\Omega_m t \right) + P(m+\kappa)\sin^2\left(\Omega_{m+\kappa} t \right)\right]\\
&+p_\downarrow\left[P(m)\cos^2\left(\Omega_{m+\kappa} t \right) + P(m-\kappa)\sin^2\left(\Omega_m t \right)\right]
\end{aligned}
\end{equation}
where $p_\uparrow=1/\left(1+e^{-\lambda_s}\right)$, $\Omega_m=\Omega\eta^\kappa \sqrt{\frac{m!}{(m-\kappa)!}}f^{(m-\kappa)}_\kappa$, and $p_\downarrow=1-p_\uparrow$. The unitless spin labour from Eq.\eqref{eq: Spin labor} is then
\begin{equation}
\begin{aligned}
    \tilde{W}_{s}&=(\braket{ J_z(t)}-\left<J_z(0)\right>)/\hbar\\
    &=\sum_m p_\downarrow P(m)\sin^2\left(\Omega_{m+\kappa} t \right)\\&-\sum_{m\ge\kappa}p_\uparrow P(m)\sin^2\left(\Omega_m t\right).\label{eq: Unitless spin labour}
\end{aligned}
\end{equation}
The average motional quanta $\bar{n}(t)=\sum_m mP(m, t)$ with a non-finite spin temperature is
\begin{equation}
\begin{aligned}
\bar{n}=p_\uparrow \sum_{m}&m\left[P(m)\cos^2\left(\Omega_m t\right)\right.\\&+\left.P(m+\kappa)\sin^2(\Omega_{m+\kappa}t)\right]\\
+p_\downarrow\sum_m& m \left[P(m)\cos^2(\Omega_{m+\kappa} t)\right.\\&+\left.P(m-\kappa)\sin^2(\Omega_m t)\right]
\end{aligned}
\end{equation}
Making the unitless work from Eq.\eqref{eq: Work as function of average quanta}
\begin{equation}\label{eq: Unitless work}
\begin{aligned}
\tilde{W}_v&=(\left<\Delta E(t)\right>-\left<E(0)\right>)/h\nu\kappa\\&=\sum_{m}p_\downarrow P(m)\sin^2(\Omega_{m+\kappa} t)\\&-\sum_{m\ge\kappa}p_\uparrow P(m)\sin^2(\Omega_m t).
\end{aligned}
\end{equation}
$\tilde{W}_{v}$ and $\tilde{W}_s$ are equal.

\bibliography{apssamp.bib}% Produces the bibliography via BibTeX.

@article{ratschbacher2013decoherence,
  title={Decoherence of a single-ion qubit immersed in a spin-polarized atomic bath},
  author={Ratschbacher, L and Sias, Carlo and Carcagni, L and Silver, JM and Zipkes, C and K{\"o}hl, M},
  journal={Physical review letters},
  volume={110},
  number={16},
  pages={160402},
  year={2013},
  publisher={APS}
}

@article{Pozsgay_2017,
doi = {10.1088/1742-5468/aa82c1},
url = {https://doi.org/10.1088/1742-5468/aa82c1},
year = {2017},
month = {sep},
publisher = {IOP Publishing and SISSA},
volume = {2017},
number = {9},
pages = {093103},
author = {Pozsgay, B and Vernier, E and Werner, M A},
title = {On generalized Gibbs ensembles with an infinite set of conserved charges},
journal = {Journal of Statistical Mechanics: Theory and Experiment}
}

@article{jaynes1957information,
  title={Information theory and statistical mechanics},
  author={Jaynes, Edwin T},
  journal={Physical review},
  volume={106},
  number={4},
  pages={620},
  year={1957},
  publisher={APS}
}

@article{janes1982rationale,
  title={On the rationale of maximum-entropy method},
  author={Jaynes, ET},
  journal={Proc. IEEE},
  volume={70},
  pages={939},
  year={1982}
}

@article{jaynes1957information2,
  title={Information theory and statistical mechanics. II},
  author={Jaynes, Edwin T},
  journal={Physical review},
  volume={108},
  number={2},
  pages={171},
  year={1957},
  publisher={APS}
}

@article{PhysRevE.101.042117,
  title = {Noncommuting conserved charges in quantum many-body thermalization},
  author = {Yunger Halpern, Nicole and Beverland, Michael E. and Kalev, Amir},
  journal = {Phys. Rev. E},
  volume = {101},
  issue = {4},
  pages = {042117},
  numpages = {13},
  year = {2020},
  month = {Apr},
  publisher = {American Physical Society},
  doi = {10.1103/PhysRevE.101.042117},
  url = {https://link.aps.org/doi/10.1103/PhysRevE.101.042117}
}

@article{Langen_2016,
doi = {10.1088/1742-5468/2016/06/064009},
url = {https://doi.org/10.1088/1742-5468/2016/06/064009},
year = {2016},
month = {jun},
publisher = {IOP Publishing and SISSA},
volume = {2016},
number = {6},
pages = {064009},
author = {Langen, Tim and Gasenzer, Thomas and Schmiedmayer, Jörg},
title = {Prethermalization and universal dynamics in near-integrable quantum systems},
journal = {Journal of Statistical Mechanics: Theory and Experiment}
}

@article{Guryanova_2016,
   title={Thermodynamics of quantum systems with multiple conserved quantities},
   volume={7},
   ISSN={2041-1723},
   url={http://dx.doi.org/10.1038/ncomms12049},
   DOI={10.1038/ncomms12049},
   number={1},
   journal={Nature Communications},
   publisher={Springer Science and Business Media LLC},
   author={Guryanova, Yelena and Popescu, Sandu and Short, Anthony J. and Silva, Ralph and Skrzypczyk, Paul},
   year={2016},
   month=jul }

@article{Vaccaro_2011,
   title={Information erasure without an energy cost},
   volume={467},
   ISSN={1471-2946},
   url={http://dx.doi.org/10.1098/rspa.2010.0577},
   DOI={10.1098/rspa.2010.0577},
   number={2130},
   journal={Proceedings of the Royal Society A: Mathematical, Physical and Engineering Sciences},
   publisher={The Royal Society},
   author={Vaccaro, Joan A. and Barnett, Stephen M.},
   year={2011},
   month=jan, pages={1770–1778} }

@article{lindblad1976generators,
  title={On the generators of quantum dynamical semigroups},
  author={Lindblad, Goran},
  journal={Communications in mathematical physics},
  volume={48},
  number={2},
  pages={119--130},
  year={1976},
  publisher={Springer}
}

@article{wright2018quantum,
  title={Quantum heat engine operating between thermal and spin reservoirs},
  author={Wright, Jackson SST and Gould, Tim and Carvalho, Andr{\'e} RR and Bedkihal, Salil and Vaccaro, Joan A},
  journal={Physical Review A},
  volume={97},
  number={5},
  pages={052104},
  year={2018},
  publisher={APS}
}

@article{PhysRevA.53.2501,
  title = {Adiabatic processes in three-level systems},
  author = {Laine, Timo A. and Stenholm, Stig},
  journal = {Phys. Rev. A},
  volume = {53},
  issue = {4},
  pages = {2501--2512},
  numpages = {0},
  year = {1996},
  month = {Apr},
  publisher = {American Physical Society},
  doi = {10.1103/PhysRevA.53.2501},
  url = {https://link.aps.org/doi/10.1103/PhysRevA.53.2501}
}

@article{mcclelland2025beyond,
  title={Beyond the Carnot limit: work extraction via an entropy battery},
  author={McClelland, Liam Judd},
  journal={arXiv preprint arXiv:2510.08989},
  year={2025}
}

@article{croucher2017discrete,
  title={Discrete fluctuations in memory erasure without energy cost},
  author={Croucher, Toshio and Bedkihal, Salil and Vaccaro, Joan A},
  journal={Physical review letters},
  volume={118},
  number={6},
  pages={060602},
  year={2017},
  publisher={APS}
}

\end{document}